\documentclass[useAMS,usenatbib]{mn2e}

\usepackage{amssymb}
\usepackage{graphicx}

\title[Radio quiet black hole binaries]{On the nature of the ``radio quiet'' black hole binaries}
\author[Paolo Soleri \& Rob Fender]{Paolo Soleri$^{1,2}$\thanks{E-mail:
soleri@astro.rug.nl} and Rob Fender$^{3,2}$\\
$^{1}$Kapteyn Astronomical Institute, University of Groningen, PO Box 800, 9700 AV, Groningen, The Netherlands\\
$^{2}$Astronomical Institute Anton Pannekoek, University of Amsterdam, Science Park 904, 1098 XH, Amsterdam, The Netherlands\\
$^{3}$School of Physics and Astronomy, University of Southampton, Hampshire, SO17 1BJ, UK}
\begin{document}

\date{Accepted 2011 January 5.  Received 2011 January 5 ; in original form 2010 June 30}

\pagerange{\pageref{firstpage}--\pageref{lastpage}} \pubyear{yyyy}

\maketitle

\label{firstpage}

\begin{abstract}
The coupling between accretion processes and ejection mechanisms in accreting black holes in binary systems can be investigated by empirical relations
between the X-ray/radio and X-ray/optical-infrared luminosities. These correlations are valid over several orders of magnitude and were initially
thought to be universal. However, recently many black hole binaries have been found to produce jets that, given certain accretion-powered luminosities, are fainter than
expected from the earlier correlations. This shows that black holes with similar accretion flows can produce a broad range of outflows in power, suggesting
that some other parameters or factors might be tuning the accretion/ejection coupling. Recent work has already shown that this jet power does not correlate with the
reported black hole spin measurements. Here we discuss whether fixed parameters
of the binary system (orbital period, disc size, inclination), as well as the properties of the outburst, produce any effect on the energy output in the jet. No obvious
dependence is found. We also show that there is no systematic variation of the slope of the radio:X-ray correlation with normalization. We define a jet-toy model in which
the bulk Lorentz factor becomes larger than $\sim 1$ above $\sim 0.1\%$ of the
Eddington luminosity. With this model, if we assume random inclination angles which result in highly variable boosting at large Eddington ratios, we are able to reproduce
qualitatively the scatter of the X-ray/radio correlation and the ``radio quiet'' population. However the model seems to be at odds with some other observed properties of
the systems. We also compare the ``radio quiet'' black holes with the neutron stars. We show that if a mass correction from the fundamental plane is applied, the
possibility that they are statistically indistinguishable in the X-ray:radio plane can not be completely ruled out. This result suggests that some of the outliers
could actually be neutron stars, or that the disc-jet coupling in the ``radio quiet'' black holes is more similar to the one in neutron stars.
\end{abstract}


\section{Introduction}
Relativistic ejections (jets) are a common consequence of accretion processes onto black holes in active galactic nuclei (AGN) as well as onto stellar mass black
holes in X-ray binaries (XRBs, see Fender 2010 for a review). In the low/hard state (LHS) and probably in the quiescent state of black hole candidates (BHCs) a
compact, steady jet is present (Fender 2001; Gallo et al. 2006). The characteristic signature of compact-steady jets (Blandford \& K\"{o}nigl 1979) is a flat/slightly inverted
spectrum ($\alpha \gtrsim 0$, $F_{\nu} \propto \nu^{\alpha}$) observed in the radio band and sometimes extending to infrared (IR) and possibly optical
frequencies (e.g. Hynes et al. 2000; Brocksopp et al. 2001). The jet power dominates over the accretion-powered luminosity in the LHS at $L_X \lesssim 1\% \, L_{Edd}$
($L_X$ and $L_{Edd}$ are the X-ray and the Eddington luminosities, respectively; Fender, Gallo \& Jonker 2003, Migliari \& Fender 2006).
There is strong evidence that the jet is highly quenched in the high/soft state (HSS) of BHCs (Tananbaum et al. 1972; Fender et al. 1999).

K\"{o}rding, Jester \& Fender (2006) showed that a generalization of the accretion states used to describe BHCs could also be applied to AGN. This suggests that
despite the different masses involved, systems that contain a black hole display similar accretion states and jet properties.

Hannikainen et al. (1998) and Corbel et al. (2003) found that the radio flux of the BHC GX 339-4 in the LHS correlates over several orders of magnitude with the X-ray flux.
Gallo, Fender \& Pooley (2003) included other sources in the sample and proposed that a correlation of the form $L_X \propto L_{R}^{b}$ (where $L_R$ is the radio luminosity) with $b = 0.58\pm0.16$ (Gallo et al. 2006) could be universal and also valid for sources in quiescence. This indicated that the
mechanisms responsible for the ejection of the outflows are closely coupled to the properties of the accretion flow. Russell et al. (2006) verified that an empirical correlation
between the X-ray luminosity and the optical/IR luminosities also holds ($L_X \propto L_{OIR}^{0.6}$, where $L_{OIR}$ is the optical/IR luminosity) for BHCs in the LHS and in
quiescence. There is evidence that in most cases the near-IR emission is jet-dominated while the optical emission is not dominated by the jet but by the reprocessing of the X-rays in the outer regions of the accretion disc (Russell et al. 2006).
\begin{figure*}
\begin{tabular}{c}
\resizebox{15cm}{!}{\includegraphics{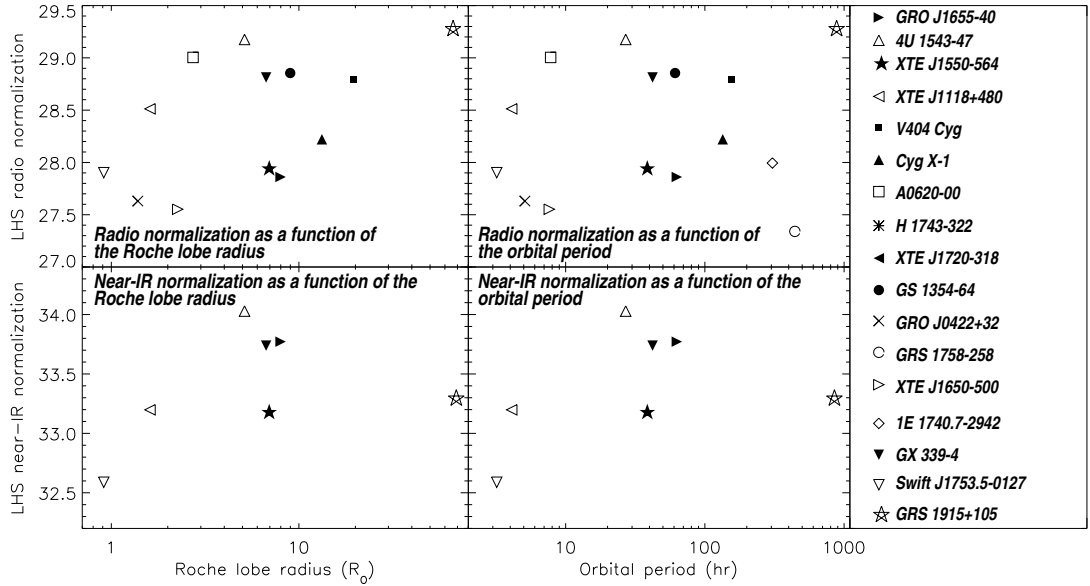}}
\end{tabular}
\caption{Radio and near-IR normalizations as a function of the orbital period and the size of the Roche lobe of the black hole. See Table \ref{tab:log_sources} for a
log of the used values. A key of the symbols used in this plot and in the following ones is in the inset.}
\label{fig:powerVSorbital_param}
\end{figure*}

Merloni, Heinz \& Di Matteo (2003) and Falcke, K\"{o}rding \& Markoff (2004) have independently shown that the same X-ray/radio scaling found for BHCs also holds for
supermassive black holes in AGN, if the mass of the compact object is taken into account. This suggests that similar mechanisms couple accretion and ejection processes to/from
black holes hold over $\ga$ 9 orders of magnitude in mass.

The existence of the radio/X-ray correlation has broad implications. For example, the small scatter around it has been used as an argument by Heinz \& Merloni
(2004) to infer that jets from BHCs and AGN (once a mass-correction factor is introduced) are characterized by similar bulk velocities, unless they are all non
relativistic. 
\begin{table*}
\centering
\caption{Radio and IR normalizations for a sample of 17 BHCs. Unless it has been differently specified, the data used to calculate the radio normalizations
are from Gallo et al. (2003, 2006) and Gallo (2007), always considering an observing frequency of 8.5 GHz. The IR normalizations are from the BHCs sample of Russell et al. (2006, 2007). We also report the orbital period
($P_{orb}$), the mass of the accretor ($M_X$), the q ratio ($q=M_X/M_2$, $M_2$ is the mass of the companion star), the size of the Roche lobe of the compact object
($R_l$), the orbital inclination (i) and the distance to the source (D). Unless more recent estimates are available, the
inclinations are from Charles \& Coe (2006) and all the other parameters from Russell et al. (2006).}
\label{tab:log_sources}
\begin{tabular}{l c c c c c c c r}
\hline
\hline
\multicolumn{9}{c}{{\bf BHC sample}}\\
            & \multicolumn{2}{c}{Normalizations} &                  \multicolumn{5}{c}{Binary properties}                              &                                \\    
Source      &    radio       &     near IR       &  $P_{orb}$ (hr)  & $M_X$($M_{\odot})$   &	  q	  & $R_l$($R_{\odot}$)         &    i ($^{\circ}$)  & D (kpc)   \\
\hline
H 1743-322  &    28.16(1,2)  &      -            &       -          &          -	   &	  -	  &	   -       &	 -	                &  7.5(1)         \\
GX 339-4    &    28.81       &    33.74          &      42.1        &       $\sim$5.8	   & $\sim12.5$   &	  6.69     & $15-60$(3)               & $8^{+7}_{-1}$   \\
XTE J1118+480 &  28.51       &    33.20          &      4.1         &       $6.8\pm0.4$    &	27.2	  &	  1.61     & $81\pm2$			 &  $1.71\pm0.05$ \\  
GRS 1915+105  &  29.26       &    33.28          &      846         &      $14.0\pm4.4$    &	17.28	  &	  64.25    & $70\pm2$			 &  $11.2\pm0.8$(3)\\
V404 Cyg      &  28.80       &      -            &      155.3       &      $10.0\pm2.0$    &	15.38	  &	  19.55    & $55\pm4$			& $2.39\pm0.14$(4)\\
A0620-00      &  29.00       &      -            &      7.75        &      $11.0\pm1.9$    &	14.86     &       2.73     & $37\pm5$                   & $1.2\pm0.4$     \\  
GRO J0422+32  &  27.63       &      -            &      5.09        &      $3.97\pm0.95$   &	8.63	  &	  1.39     & $45\pm2$                   & $2.49\pm0.30$   \\
GS 1354-64    &  28.85       &      -            &      61.1(5)     &    $7.83\pm0.50$(5)  &    7.68(5)   &       8.99     & $<79$                      & $>25$(5)        \\
4U 1543-47    &  29.18       &    34.03(6)       &      26.8        &      $9.4\pm1.0$     &    3.84      &       5.13	   & $21\pm2$                   & $7.5\pm0.5$     \\
XTE J1550-564 &  27.92       &    33.16(6)       &      36.96       &      $10.6\pm1.0$    &	7.41      &       7.16	   & $72\pm5$                   & $4.1\pm0.8$(7)  \\
GRO J1655-40  &  27.86(8)    &    33.77          &      62.9        &      $7.02\pm0.22$   &    2.99      &       7.98     & $70\pm2$                   & $\leq 1.7$(9)  \\
XTE J1650-500 &  27.55       &      -            &      7.63        &      $<7.3$(10)      &  $>10$(10)   &       2.26     & $50\pm3$(10)               & $2.6\pm0.7$(11) \\   
Swift J1753.5-0127 & 27.90(12) &  32.59(12)      &      3.2(13)     &      $>3.0$(13)   & $\gtrsim10$(13) &       0.91     & $>85$(14)                  & $\sim 8$(13,15) \\
Cyg X-1       &   28.22      &      -            &      134.4(16)   &   $\sim10.1$(16)     &    0.57(16)  &       13.26    & $35\pm5$(3)                & $\sim 2.1$      \\
XTE J1720-318 &   27.49      &      -            &       -          &          -	   &	  -	  &	   -       &	 -	                & $6.5\pm3.5$(17) \\   
1E1740.7-2942 &   27.99      &      -            &      305.52(18)  &          -	   &	  -	  &	   -       &     -                      & $\sim 8.5$(3)   \\
GRS 1758-258  &   27.34      &      -            &      442.8(18)   &          -	   &	  -	  &	   -       &     -                      & $\sim 8.5$(3)   \\
\hline  
\hline
\end{tabular}
\small
\\ (1) Jonker et al. (2010); (2) McClintock et al. (2009); (3) Gallo et al. (2003); (4) Miller-Jones et al. (2009); (5) Casares et al. (2009);
(6) Russell et al. (2007); (7) Hannikainen et al. (2009); (8) Migliari et al. (2007a); (9) Foellmi et al. (2006); (10) Orosz et al. (2004);
(11) Homan et al. (2006); (12) Soleri et al. (2010); (13) Zurita et al. (2008); (14) Hiemstra et al. (2009); (15) Cadolle Bel et al. (2007);
(16) Herrero et al. (1995); (17) Chaty \& Bessolaz (2006); (18) Smith et al. (2002)
\normalsize
\end{table*}

However, in the past few years, the supposed universality of the radio/X-ray correlation has been doubted (Xue \& Cui 2007) and several ``radio quiet''
outliers have been found (Gallo 2007). These sources seem to feature similar X-ray luminosities to other
BHCs but are characterized by a radio emission that, given a certain X-ray luminosity, is fainter than
expected from the radio/X-ray correlation. It is possible that a correlation with similar slope but lower normalization than the other BHCs could describe this
discrepancy, at least in a few sources (e.g. Corbel et al. 2004; Gallo 2007; Soleri et al. 2010). 
If confirmed, this would suggest that some other parameters might be tuning the accretion-ejection coupling, allowing accretion flows with similar radiative
efficiency to produce a broad range of outflows.

Garcia et al. (2003) investigated the dependence of the jet power on the orbital period of the binary. They noted that, among 14 dynamically confirmed BHCs, we can
spatially resolve a powerful jet in 4 systems characterized by long orbital periods. Although these radio jets are most likely blobs launched during major ejection events (and not compact-steady jets typical of BHCs in the LHS, see Fender, Belloni \& Gallo 2004), this could suggest that the orbital period might play a role in powering jets from
BHCs.
\begin{figure}
\begin{tabular}{c}
\resizebox{8.5cm}{!}{\includegraphics{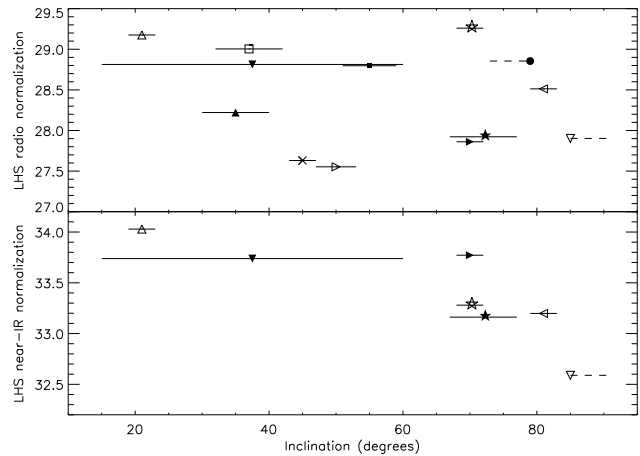}}
\end{tabular}
\caption{Radio and near-IR normalizations as a function of the inclination angles $i$ of the BHCs. See Table \ref{tab:log_sources} for a list of the used values. The dashed
lines indicate an upper limit (GS~1354-64, filled circle) and a lower limit (Swift~J1753.5-0127, downward open triangle). See Figure \ref{fig:powerVSorbital_param} for a key
to the symbols.}
\label{fig:powerVSinclination}
\end{figure}

Pe'er \& Casella (2009) presented a model for the emission from jets in XRBs in which the electrons are accelerated only once at the base of
the jet (at variance with other models, in which multiple accelerations occur; see e.g. Maitra et al. 2009, Jamil, Fender \& Kaiser 2009). In the model, the jet magnetic field
is a parameter that can cause a quenching of the radio emission (when above a critical value $B_{cr} \approx 10^{5}\, \mbox{G}$), without influencing the energy output in the X rays (Casella \& Pe'er 2009).\\ 
The dependence of the jet power on the spin of the black hole has recently been investigated by Fender, Gallo \& Russell (2010). They inferred the jet power from the
normalizations of the radio/X-ray and near-IR/X-ray correlations found by Gallo et al. (2006) and Russell et al. (2006). They concluded that, if our measures of the spin and the
estimates of the jet power are correct, the spin does not play any role in powering jets from BHCs. In AGN, on the other hand, the most powerful jets have been associated with high-spin black holes (Sikora, Stawarz \& Lasota 2007).

In this paper we investigate whether there is any connection between the values of some fixed binary parameters and properties of the outburst of BHCs and the compact
steady-jet power. We discuss how Doppler de-boosting effects could
qualitatively explain the scatter around the radio/X-ray correlation, when a particular dependence of the bulk Lorentz factor of the jet on the accretion-powered
luminosity is taken into account.
We also compare the ``radio quiet'' BHCs to the neutron star (NS) XRBs, since these systems are usually fainter in radio (given a certain X-ray luminosity) than BHCs (see e.g. Migliari \& Fender 2006).

\section{The hard state jet power} \label{par:power}
In this paper we will follow the approach presented in Fender et al. (2010) to use the normalizations of the radio/X-ray
and OIR/X-ray correlations as a proxy for the jet power. We will consider the slopes of the correlations
as $\sim 0.6$. Although some sources have been found to follow
a correlation with a different slope (e.g. H 1743-322, $b \sim 0.18$; Jonker et al. 2010)
and this parameter is often badly constrained by fitting the data points (e.g. Swift J1753.5-0127, Soleri et al. 2010), we
can consider it universal (see \S \ref{par:discussion} for a discussion on the applicability of this method), in order to have a rough estimate of the jet power. For more details on this method we refer the reader to Fender et al. (2010).

Table \ref{tab:log_sources} lists the BHCs considered in this paper. Unless otherwise specified, the normalizations have been calculated using the data from Gallo et
al. (2003, 2006), Gallo (2007) and Russell et al. (2006, 2007). For the radio data, we always considered an observing frequency of 8.5 GHz. Since the optical emission is not dominated by the jet, we used only the near-IR data from Russell et
al. (2006, 2007; J, K and H bands). For each source we fitted the X-ray and radio data using a relation of the form $log_{10}(L_{R}) = c_R + 0.6(log_{10}(L_X) - 34)$, considering the normalizations $c_R$ as free parameters. Since the X-ray and radio data that we fitted do not
have an error, we considered the root mean square (rms) of the residuals (the dispersion of the data points around the
best fit relation) as an estimate of the uncertainty of $c_R$. Sources that cannot be well fitted using a slope $b \sim 0.6$
will give high values of the rms of the residuals. We applied the same method to obtain the normalizations
$c_{IR}$ and the rms of the residuals from the X-ray and near-IR data.
Since Gallo et al. (2003) and Russell et al. (2006) in some cases adopted different distances to the
same source, in this paper we use the most recent estimates. For the BHC Cyg X-1 we do not include the data points that show
evidence for suppression of the radio emission as the source enters softer X-ray states (see Figure 3 of Gallo et al. 2003
for the details). Although we are only considering BHCs in the LHS, we include in our
sample GRS 1915+105 (which spends all its time in the intermediate states), using data from the radio-bright
plateau state, which is approximately an analogous to the LHS (see Fender \& Belloni 2004 for a review on the source).
The properties of the BHCs are reported in Table \ref{tab:log_sources}, as well as the normalizations. 
($c_{R}$ and $c_{IR}$) used as a proxy for the jet power. The rms of the residuals and the number of data points used for each
fit are reported in Table \ref{tab:rms_residuals}, while
the properties of the outbursts of our BHCs are listed in Table \ref{tab:log_outbursts}.
\begin{table}
\centering
\caption{rms of the residuals obtained fitting the X-ray and radio data with a relation of the form $log_{10}(L_{R}) = c_R + 0.6(log_{10}(L_X) - 34)$, leaving $c_R$ as a free parameter. The same has been done for the near-IR and the X-ray data.
The table also shows the number of data points used for each fit. For some BHCs
we only had one data point, hence the rms of the residuals could not be calculated. Instead, for these
sources, we report the average of the rms of the other systems.}
\label{tab:rms_residuals}
\begin{tabular}{l c c c c}
\hline
\hline
                   &  \multicolumn{2}{c}{rms of residuals} & \multicolumn{2}{c}{Number of points} \\
Source             & radio            & near IR            &  radio &   near IR \\
\hline  
H 1743-322         &   0.63           &      -             &  14    &  -        \\
GX 339-4           &   0.12           &   0.09             &  12    &  23       \\
XTE J1118+480      &   0.08           &   0.14             &  34    &  14       \\  
GRS 1915+105       &   0.20           &   0.12             &   1    &  1        \\
V404 Cyg           &   0.11           &     -              &  21    &  -        \\
A0620-00           &   0.20           &     -              &   1    &  -        \\  
GRO J0422+32       &   0.16           &     -              &   2    &  -        \\
GS 1354-64         &   0.06           &     -              &   3    &  -        \\
4U 1543-47         &   0.20           &   0.05             &   1    &  17       \\
XTE J1550-564      &   0.20           &   0.28             &   1    &  20       \\
GRO J1655-40       &   0.51           &   0.12             &   2    &  1        \\
XTE J1650-500      &   0.18           &     -              &   4    &  -        \\   
Swift J1753.5-0127 &   0.20           &   0.02             &  30    &  9        \\
Cyg X-1            &   0.12           &    -               &  1029  &  -        \\
XTE J1720-318      &   0.06           &    -               &   2    &  -        \\   
1E1740.7-2942      &   0.20           &    -               &   1    &  -        \\
GRS 1758-258       &   0.20           &    -               &   1    &  -        \\
\hline  
\hline
\end{tabular}
\end{table}

\section{BHC properties and jet power} \label{par:jet_power}
We will now examine whether three characteristic parameters of the binary system (the orbital period, the size of the accretion disc and the orbital inclination)
and the properties of the outburst affect the energy output in the jet.
\begin{table}
\centering
\caption{Properties of the outbursts of 10 BHCs in our sample. The sources that showed only ``normal'' outbursts are not listed here. See Brocksopp et al. (2004) for the
details. HIMS and SIMS mean hard-intermediate state and soft-intermediate state, respectively (see Belloni 2009 for a definition of the states).}
\label{tab:log_outbursts}
\begin{tabular}{l l l}
\hline
\hline
Source        & Outbursts occurred                 & Additional remarks                \\
\hline  
H 1743-322    & several, normal                    &  we use data                      \\
              & or LHS-HIMS                        &  from a normal one                \\
              & only(1)                            &                                   \\
GRS 1915+105  & quasi persistent,	           &  we use data from                 \\	
	      & in the  	                   &  the radio-bright	               \\
	      & HIMS-SIMS(2)                       &  plateau state                    \\
V404 Cyg      & one, LHS only			   &  possible transition              \\
              &                                    &  to the HSS(3)                    \\
XTE J1118+480 & two, LHS only			   &  we use data from  	       \\
              &	                                   &  the first one                    \\
GRO J0422+32  & one, LHS only			   & \multicolumn{1}{c}{-}	       \\   
GS 1354-64    & four, normal                       &  we use data from                 \\ 
              & and LHS only                       &  a LHS-only one                   \\
4U 1543-47    & four, normal                 	   &  we use data from                 \\
              & and LHS only			   &  a normal one                     \\
XTE J1550-564 & four, normal	                   &  we use data from                 \\
              & and LHS only                       &  a normal one                     \\
Swift J1753.5-0127 & one (ongoing),		   &  see (5), possible  	       \\  
              & LHS only(4)                        &  transition to the                \\
              &                                    &  intermediate states              \\
Cyg X-1       & persistent			   &  see (6) for details              \\
              &                                    &  on the data                      \\
              &                                    &  selection	                       \\
\hline  
\hline
\end{tabular}
\small
\\ (1) Capitanio et al. (2009); (2) Fender \& Belloni (2004); (3) \citet*{Zycki1999}; (4) Cadolle Bel et al. (2007); (5) Negoro et al. (2009);
(6) Gallo et al. (2003)
\normalsize
\end{table}

\subsection{Binary parameters} \label{par:binary}
\begin{figure}
\begin{tabular}{c}
\resizebox{8.5cm}{!}{\includegraphics{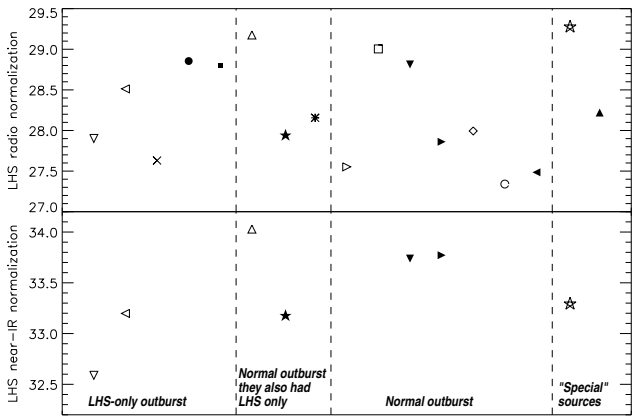}}
\end{tabular}
\caption{Radio and near-IR normalizations as a function of the properties of the outburst from which they have been obtained. By ``normal'' we mean an outburst that shows
a transition to the soft states. The ``special'' sources are GRS 1915+105 and Cyg X-1, which are a quasi-persistent and a persistent source, respectively. See Figure
\ref{fig:powerVSorbital_param} for a key to the symbols.}
\label{fig:power_outburst}
\end{figure}
Since the accretion disc occupies $\sim 70\%$ of the Roche lobe of the black hole (Frank, King \& Raine 2002), we calculated the size of the Roche lobe of the accretor as a
measure of the disc size. Following Frank, King \& Raine (2002), the orbital separation $a$ is given by:
$$a= 3.5 \times 10^{10}M_{X}^{1/3}(1+\frac{1}{q})^{1/3} \, P_{orb}^{2/3} \, \mbox{cm}$$
where $M_X$ is the mass of the accretor (in $M_{\odot}$ units), $q$ is the mass ratio ($q = M_{X}/M_{2}$, $M_2$ is the mass of the donor) and $P_{orb}$ the orbital
period (in hours). The size of the Roche lobe of the compact object $R_{l}$ can be calculated as follows: 
$$R_l = a\, \frac{0.49 \, q^{2/3}}{0.6 \,q^{2/3} + \mbox{ln}(1+q^{1/3})}\, \mbox{cm}$$
Figure \ref{fig:powerVSorbital_param} shows the radio and near-IR normalizations as a function of the size of the Roche lobe of the black hole and the orbital period of the
binary. The lower panels of Figure \ref{fig:powerVSorbital_param} might suggest that the near-IR normalization increases with the size of the Roche lobe of the accretor and
with the orbital period, although the sample contains only 7 BHCs. Two of them do not follow this possible trend: XTE~J1550-564 and GRS~1915+105.

To test whether there is any correlation between the jet power and these two orbital parameters, we calculated the Spearman rank correlation coefficients for the data points in Figure \ref{fig:powerVSorbital_param}. The values of the correlation coefficient $\rho$, as well as the null hypothesis probabilities (the probability that the data are not correlated), are reported in Table \ref{tab:log_spearman}. Clearly no correlation is present.
\begin{table*}
\centering
\caption{Spearman rank correlation coefficient for the data points in Figures \ref{fig:powerVSorbital_param} and \ref{fig:powerVSinclination}. The number of data points, as well as the probabilities for the null hypothesis, are also reported.
We also calculated the Spearman rank correlation coefficients (and the probabilities for the null hypothesis) for
a restricted sample of sources for which both mass and distance have been measured.
Considering the radio normalizations, the restricted sample includes XTE J1118+480, GRS 1915+105, V404 Cyg,
A0620-00, GRO J0422+32, 4U 1543-47, XTE J1550-564 and Cyg X-1. For the near-IR normalizations,
the sources in the restricted sample are XTE J1118+480, GRS 1915+105, 4U 1543-47 and XTE J1550-564.
We also corrected the radio luminosities for the sources in the restricted sample by applying 
the mass correction from the fundamental plane of black hole activity of Merloni et al. (2003).
The Spearman rank correlation coefficient and the probabilities of the null hypothesis are reported at the end
of the table.}
\label{tab:log_spearman}
\begin{tabular}{l c c c}
\hline
\hline
                              \multicolumn{4}{c}{{\bf Spearman rank correlation test - all the sources}}                 \\
\hline  
Normalization          & Number of points  & Spearman coefficient $\rho$   & Probability for the null hypothesis (\%)  \\
                                       \multicolumn{4}{c}{{\bf Size of the Roche lobe}}                              \\                     
radio                  &       13        &          0.5                  &                 11.0                       \\
near IR                &       7         &          0.4                  &                 37.9                       \\
                              \multicolumn{4}{c}{{\bf Orbital period}}                                                 \\
radio                  &      15         &          0.2                  &                 54.2                       \\
near IR                &      7          &          0.5                  &                 25.4                       \\
                              \multicolumn{4}{c}{{\bf Inclination angle}}                                              \\
radio                  &       13        &          -0.2                 &                 44.7                       \\
near IR                &       7         &          -0.9                 &                 3.0                        \\
\hline
\hline
               \multicolumn{4}{c}{{\bf Spearman rank correlation test - restricted sample, no mass correction}}        \\
\hline  
Normalization          & Number of points  & Spearman coefficient $\rho$   & Probability for the null hypothesis (\%)  \\                                       \multicolumn{4}{c}{{\bf Size of the Roche lobe}}                              \\                     
radio                  &       8        &          0.5                  &                 20.8                       \\
near IR                &       4        &          $\sim 0$             &                 99.2                       \\
                              \multicolumn{4}{c}{{\bf Orbital period}}                                                 \\
radio                  &      8          &          0.4                  &                 28.0                       \\
near IR                &      4          &          $\sim 0$             &                 99.2                       \\
                              \multicolumn{4}{c}{{\bf Inclination angle}}                                              \\
radio                  &       8         &          -0.2                 &                 61.0                       \\
near IR                &       4         &          -0.8                 &                 16.5                        \\
\hline
               \multicolumn{4}{c}{{\bf Spearman rank correlation test - restricted sample, mass corrected}}        \\
\hline  
Normalization          & Number of points  & Spearman coefficient $\rho$   & Probability for the null hypothesis (\%)  \\                                       \multicolumn{4}{c}{{\bf Size of the Roche lobe}}                              \\                     
radio                  &       8        &          0.3                   &                44.7                       \\
                              \multicolumn{4}{c}{{\bf Orbital period}}                                               \\
radio                  &      8          &          0.2                  &                52.9                       \\
                              \multicolumn{4}{c}{{\bf Inclination angle}}                                            \\
radio                  &       8         &         -0.4                  &                31.2                       \\
\hline  
\hline
\end{tabular}
\end{table*}

We also investigate the dependence of the radio and near-IR normalizations on the orbital inclinations of the BHCs. Although there is evidence that in some sources
the binary inclination does not coincide with the inclination between the jet axis and the line of sight (e.g. the misalignment has been
estimated to be $\sim 15^{\circ}$ in GRO J1655-40, see Maccarone 2002 and references therein), in this paper we will consider the orbital inclination as a proxy
for the jet axis inclination to the line of sight. We will refer to this angle $i$ as either inclination or viewing angle.
Compact-steady jets in the LHS are
thought to be mildly relativistic (with bulk Lorentz factor $\Gamma < 2$; Gallo et al. 2003, Fender et al. 2004). That suggests that de-boosting effects should
not be relevant and the jets, except for high inclination angles, should not be de-boosted. However, new results cast doubts on this fundamental assumption. Casella
et al (2010) recently observed the BHC GX 339-4 in the LHS at low X-ray luminosity ($L_X \sim 0.14\% \, L_{Edd}$, considering a distance to the source and a mass $M_X$ as in
Table \ref{tab:log_sources}), with coordinated X-ray and IR observations at high-time resolution. From the analysis of the cross-correlation function, they inferred a lower
limit on the bulk Lorentz factor of the jet $\Gamma >2$ (at $3.8 \sigma$ confidence level). This result suggests that de-boosting effects can become important,
not only at high viewing angles. Assuming that the X-ray emission is un-beamed, jets with $\Gamma \gtrsim 2$ and not pointing towards us, should result less luminous than expected from the empirical radio/X-ray and OIR/X-ray correlations. However, the possibility that the X rays are coming from the base of the jet can not be discarded
(see Markoff, Nowak \& Wilms 2005 for a theoretical model), since there is now evidence that, at least in the BHC XTE J1550-564, the synchrotron emission from the compact-steady jet dominates the X-ray emission (in the luminosity range $2 \times 10^{-4} - 2 \times 10^{-3}\, L_{Edd}$, Russell et al. 2010). Bearing this in mind, in this paper we will consider the X-ray emission as un-beamed.

Figure \ref{fig:powerVSinclination} shows the radio and near-IR normalizations as a function of the inclination angles $i$. From the upper panel, no obvious dependence can be
found. In the lower panel, the distribution of the data points suggests that BHCs characterized by a high inclination could have a low near-IR normalization.
To test if a correlation exists, we calculated the Spearman coefficient $\rho$ for the data points. We show them in Table \ref{tab:log_spearman}. In the case of the near-IR normalizations, we obtained $\rho \sim -0.9$, with a probability for the null hypothesis of $\sim 3 \%$. This represents marginal evidence for an anticorrelation between the inclination angle and the near-IR normalization. However, the lack of data points (compared to the upper panel of Figure \ref{fig:powerVSinclination}, in which no correlation can be found) might have biassed this result. 

\subsection{Binary parameters: a restricted sample} \label{par:binary_restricted}
In this sub-section we repeat the same analysis done in \S \ref{par:binary} on a restricted sample of sources.
We will consider only those BHCs for which both mass and distance have been measured: XTE J1118+480, GRS 1915+105,
V404 Cyg, A0620-00, GRO J0422+32, 4U 1543-47, XTE J1550-564 and Cyg X-1. Table \ref{tab:log_spearman} reports the 
Sperman correlation coefficients as well as the probabilities of the null hypothesis. The radio normalizations
are not correlated to the size of the Roche lobe, the orbital period and the orbital inclination. The near-IR
jet power does not show any correlation either. However, in the near-IR sample only 4 BHCs are present.

We also calculated new radio luminosities for the sources in the restricted sample by applying a mass correction. 
Considering the fundamental plane of black hole activity of Merloni et al. (2003), the mass corrected radio
luminosity is $L_{R,corr} = L_{R} / M^{0.78}$, where $M$ is the mass of the compact object. Using $L_{R,corr}$
we calculated the new mass-corrected radio normalizations. The Spearman
correlation coefficients are reported in Table \ref{tab:log_spearman}: the mass corrected radio normalizations
and the 3 orbital parameters examined in these paper are not correlated.

\subsection{Properties of the outburst} \label{par:outburst}
During an outburst, BHCs usually show a transition to the HSS (see Belloni 2009 and references therein). The transition to the soft states is characterized by sudden changes in the jet properties (Tananbaum et al. 1972; Fender et al. 1999) and it is possibly
associated with the emission of highly relativistic jets (with $\Gamma >2$; Fender et al. 2004).
Furthermore, Russell et al. (2007) and Corbel et al. (in preparation) found a dependence of the near-IR and radio
normalizations on the phase of the outburst in which the BHC is observed (e.g. XTE J1550-564).
More specifically, the normalizations measured at the outburst rise (in the LHS) are different from the normalizations
measured at the outburst decay, after the source has transited back from the HSS to the LHS. For these reasons, we
could imagine that even in the LHS, the jet properties (e.g. its bulk velocity and power) are influenced by the
transition to the soft states.
However, some sources spend the whole outburst in the LHS, without transiting to the soft states (Brocksopp,
Bandyopadhyay \& Fender 2004). Since it is not clear why some BHCs do not soften, it is important to ascertain whether
LHS-only outbursts feature different jet properties.

Table \ref{tab:log_outbursts} presents the properties of the outbursts for 10 BHCs in our sample.
Sources that had only ``normal'' outbursts (with ``normal'' we refer to BHCs that showed a transition to the soft states)
have not been listed.

To see whether the type of outburst (LHS only or with a transition to the soft states) affects the jet power,
in Figure \ref{fig:power_outburst} we reported the radio and near-IR normalizations of our sample of BHCs, dividing it
according to the properties of the outburst. No obvious dependence of the jet power on the
type of the outburst can be found.

\section{The scatter of the radio/X-ray correlation: a jet-toy model} \label{par:boosting}
Here we define a jet-toy model. The aim is to test whether a dependence of the bulk Lorentz factor of the jet $\Gamma$ on the accretion powered X-ray luminosity might
qualitatively describe the scatter around the radio/X-ray correlation and the ``radio quiet'' BHCs population. As mentioned above, we are assuming that the
X-ray emission is un-beamed (but see Markoff et al. 2005 and Russell et al. 2010).

\begin{figure*}
\begin{tabular}{c}
\resizebox{17.0cm}{!}{\includegraphics{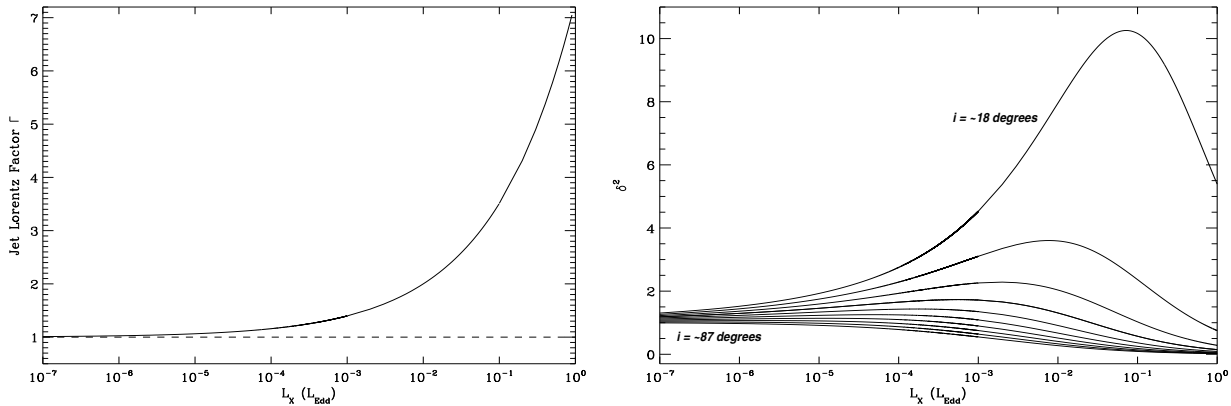}}
\end{tabular}
\caption{Left-hand panel: Lorentz factor of the jet (solid line) as a function of $L_X$ (in Eddington units, for a $\sim 10 M_{\odot}$ black hole) as used in our toy model.
$\Gamma$ becomes larger than $\sim 1.4$ above $\sim 0.1 \%$ of $L_{Edd}$. Right-hand panel: Doppler boosting factor $\delta^2$ as a function of $L_X$, for 10 possible viewing
angles (in the range $i \sim 18^{\circ}-87^{\circ}$, this corresponds to $\mbox{cos}\, i \sim 0.05 - 0.95$).}
\label{fig:de_boosting}
\end{figure*}

\begin{figure*}
\begin{tabular}{c}
\resizebox{14cm}{!}{\includegraphics{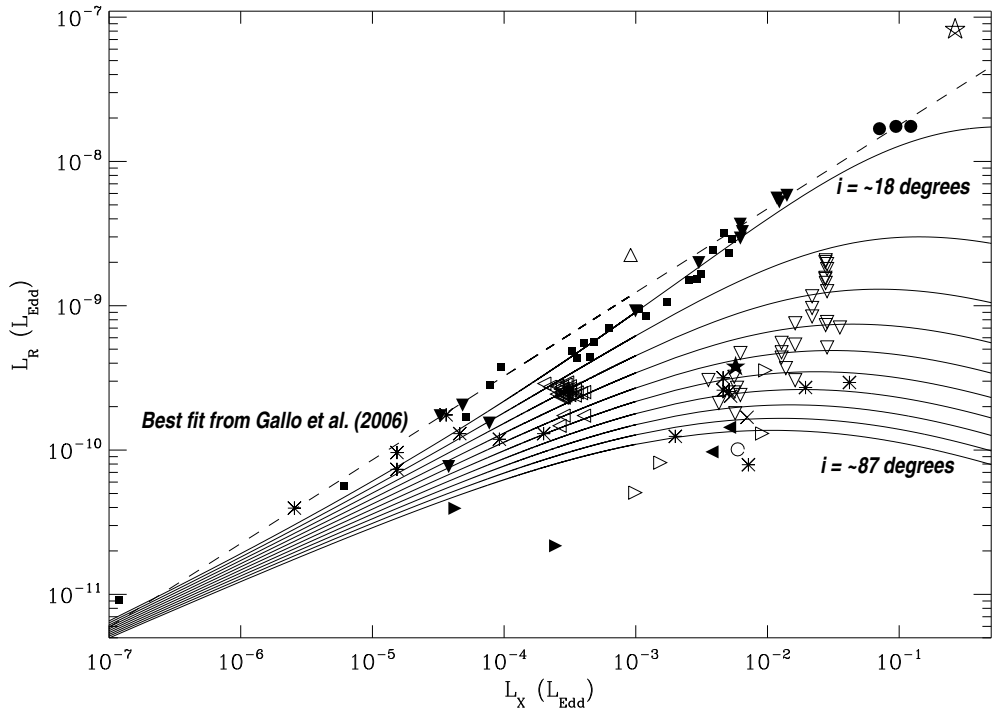}}
\end{tabular}
\caption{Values of $L_R$ expected from our toy model for 10 viewing angles (in the range $0.05 \leq \mbox{cos}\, i \leq0.95$), in Eddington units. We used the same mass for all the BHCs ($ 10 M_{\odot}$) The data points used to infer the radio
normalizations $c_{R}$ are also plotted.
For clarity, we did not include the Cyg X-1 data points. A key to the different symbols is in Figure \ref{fig:powerVSorbital_param}.
The dashed line represents the best fit correlation obtained in Gallo et al. (2006).}
\label{fig:toy_model}
\end{figure*}

We will consider a $\Gamma$ Lorentz factor that becomes larger than $\sim 1.4$ above $\sim 0.1 \%$ of the Eddington luminosity $L_{Edd}$ (see Figure \ref{fig:de_boosting},
left-hand panel). This assumption is based on the fact that compact steady jets are thought to be mildly relativistic ($\Gamma \leq 2$; Fender et al. 2004, but see Casella et al. 2010)
in the LHS below about $1\%$ of $L_{Edd}$ (e.g. Fender et al. 2003, Migliari \& Fender 2006) while major relativistic ejections ($\Gamma \geq 2$) are
tentatively associated with the transition from the hard to
the soft states. These transitions occur at a variable $L_X$ but usually above a few per cent of $L_{Edd}$ (Fender et al. 2004). We calculated the Doppler boosting factor
$\delta$ ($\delta = \Gamma^{-1} \times (1- \beta\, \mbox{cos}\, i)^{-1}$; $\beta = v/c$ is the bulk velocity of the emitting material;
$\Gamma=(1-\beta^2)^{-1/2}$) for 10 possible inclinations $i$.
Since the orientation of the approaching jet is random on a hemisphere of $2 \pi$ sr,
$\mbox{cos}\, i$ is uniformly distributed between 0 and 1. This does not imply that $i$ is uniformly distributed in the range $0^{\circ}-90^{\circ}$. We considered a
uniform distribution of 10 values of $\mbox{cos}\, i$ between 0 and 1. Figure \ref{fig:de_boosting} (right-hand panel) shows the evolution of $\delta^2$ as a function
of $L_X$. At $L_X \gtrsim 10\% \, L_{Edd}$, only for one inclination angle (of the ten considered) the jet will be boosted ($\delta^2 > 1$) and not de-boosted.
If the the jet power is well traced by
the radio luminosity $L_R$ and $L_{R} \propto L_{X}\, \delta^2$, for each inclination $i$ we can determine how the Doppler de-boosting will affect the radio-jet luminosity $L_R$. 
Figure \ref{fig:toy_model} illustrates the results from our toy model: it clearly results in a distribution in the ($L_X,L_R$) plane which broadens at higher
luminosities. The data points represents the BHCs sample used to infer the radio normalizations reported in Table \ref{tab:log_sources}. The scatter around the best-fit correlation of Gallo et al. (2006, dashed line) can be partially reproduced.
\begin{table}
\centering
\caption{NSs sample considered in this paper, from Migliari \& Fender (2006). We added data points from recent observations of Aql X-1, 4U 0614-091 and IGR J00291+5934 (see the references). 
The radio and X-ray luminosities used for Figure \ref{fig:BH_NS} have been calculated using the most recent estimates of the distances (from Migliari \& Fender 2006 if not differently specified), considering an observing radio frequency of 8.5 GHz.}
\label{tab:log_NS}
\begin{tabular}{l c c}
\hline
\hline
\multicolumn{3}{c}{{\bf Neutron stars}}                                                \\
\multicolumn{1}{c}{{\bf Source}} & {\bf Distance (kpc)} & {\bf References} \\
\hline  
                       \multicolumn{3}{c}{{\bf Atoll sources}}                         \\
4U 1728-34       & 4.6                                     &  (1)                      \\
Ser X-1          & 12.7                                    &   -                       \\ 
4U 1820          & 7.6                                     &   -                       \\ 
MXB 1730-335     & 8.8                                     &   -                       \\ 
4U 0614-091      & 3.2                                     &  (2,3)                    \\
4U 1608-52       & 5.8                                     &  (4)                      \\
                               \multicolumn{3}{c}{{\bf Accreting millisecond X-ray pulsars}}                         \\
Aql X-1          & 5.2                                     &  (5)                      \\ 
IGR J00291+5934  & 3                                       &  (6,7)                    \\
SAX J1808.4-3658 & 3.5                                     &  (8)                      \\
                               \multicolumn{3}{c}{{\bf Z sources}}                     \\
Sco X-1          & 2.8                                     &   -                       \\
GX 17+2          & 14                                      &   -                       \\
GX 349+2         & 5                                       &   -                       \\  
Cyg X-2          & 13.3                                    &   -                       \\
GX 5-1           & 9.2                                     &   -                       \\  
GX 340+0         & 11                                      &   -                       \\  
GX 13+1          & 7                                       &   -                       \\  
\hline  
\hline
\end{tabular}
\small
\\ (1) Falanga et al. (2006); (2) Migliari et al. (2010); (3) Kuulkers et al. (2009); (4) G{\"u}ver et al. (2010); (5) Tudose et al. (2009), (6) Torres et al. (2008); (7) Lewis et al. (2010); (8) Galloway \& Cumming (2006);
\normalsize
\end{table}

\section{Comparison with neutron stars} \label{par:comparison_NS}
\begin{figure*}
\begin{tabular}{c}
\resizebox{14cm}{!}{\includegraphics{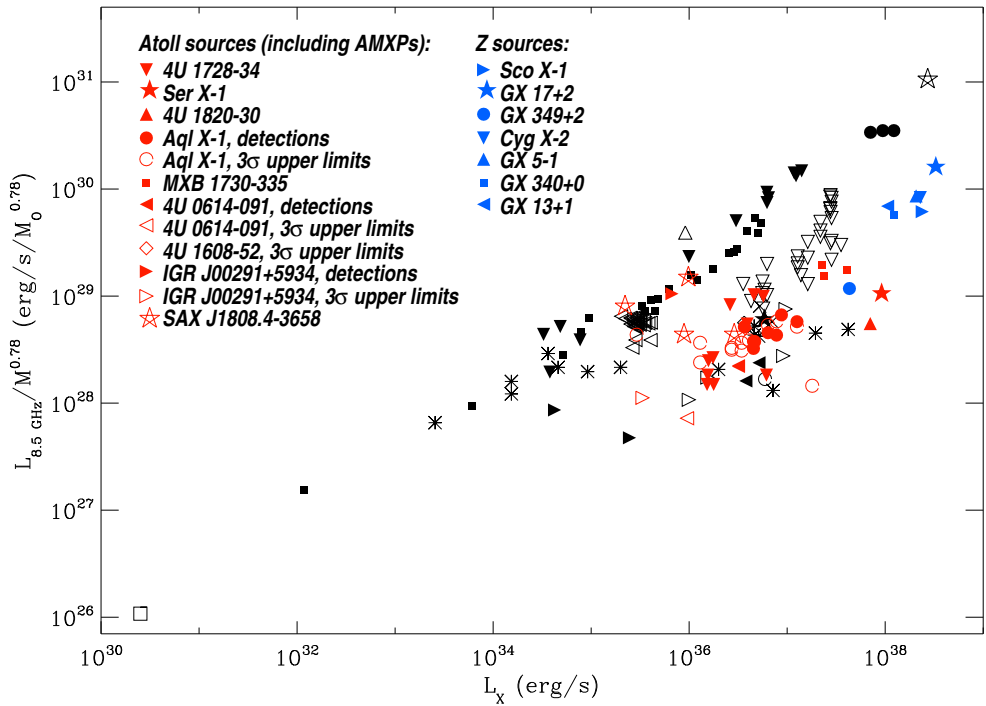}}
\end{tabular}
\caption{Our sample of BHCs (from Table \ref{tab:log_sources}), with the addition of the NSs sample from Migliari \& Fender (2006) (see Table \ref{tab:log_NS}). We also included recent data of Aql X-1, 4U 0614-091 and IGR J00291+5934, reported in Tudose et al. (2009), Migliari et al. (2010) and Lewis et al. (2010), respectively (see Table \ref{tab:log_NS}). Following Merloni et al. (2003), we applied a mass correction (considering $M_{NS} \approx 1.4 M_{\odot}$ and the black hole masses in Table
\ref{tab:log_sources}). The key to the BHC symbols is in Figure \ref{fig:powerVSorbital_param}. We excluded the BHC Cyg X-1, for clarity.}
\label{fig:BH_NS}
\end{figure*}

NSs are known to be fainter in radio than BHCs, given a certain X-ray luminosity, by a factor $\gtrsim 30$ (see Migliari \& Fender 2006 and references therein). This difference in radio power can be reduced to a factor $\gtrsim 7$ if a mass correction
from the fundamental plane of black hole activity of Merloni et al. (2003) is applied (see \S \ref{par:binary_restricted}).
We will now compare the ``radio quiet'' BHCs to the population of NSs that have been detected in radio. Our sample of NSs is presented in Table \ref{tab:log_NS}: we considered the same data points as in Migliari \& Fender (2006) with the addition of points from recent observations of Aql~X-1, 4U~0614-091 and IGR~J00291+5934. The luminosities have been calculated using the latest estimates of the distances. Figure \ref{fig:BH_NS} shows the mass corrected radio luminosity for our sample of BHCs and NSs as a function of the X-ray luminosity. The NS points seem to overlap to the ``radio quiet'' BHC points. To test whether these two groups of data points are statistically distinguishable, we performed a two-dimensional Kolmogorov-Smirnov (K-S) test (Peacock 1983; Fasano \& Fanceschini 1987), excluding all the upper limits. Our results are reported in Table \ref{tab:KS}.
The K-S test shows that the probability that the ``radio quiet'' BHCs and the NSs are statistically indistinguishable (i.e. the probability of the null hypothesis) is different from 0, despite being small ($P \sim 0.13 \%$). If we do not apply a mass correction, the probability of the null hypothesis is consistent with 0: the two groups constitute two different populations.
The two-dimensional K-S test gives similar results considering only the atoll sources (including the accreting millisecond X-ray pulsars), instead of all the NSs, both applying or not a mass correction factor.

\section{Discussion} \label{par:discussion}
\begin{figure*}
\begin{tabular}{c}
\resizebox{14cm}{!}{\includegraphics{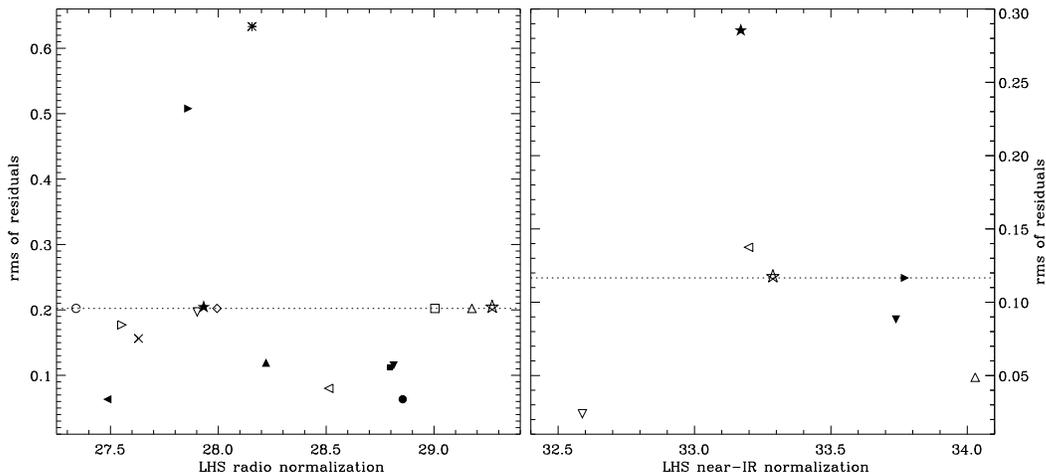}}
\end{tabular}
\caption{Rms of the residuals versus the radio and near-IR normalizations (right-hand and left-hand panel, respectively). For some BHCs
we only had one data point (see Table \ref{tab:rms_residuals}), hence we could not calculate the rms of the residuals. For them we used the average of the rms of the residuals of the other sources. The average rms is marked with a dotted line, in both panels. A key to the different symbols
is in Figure \ref{fig:powerVSorbital_param}.}
\label{fig:normalizations_VS_rms}
\end{figure*}

In this paper we tested whether there is a connection between the values of some characteristic binary parameters of BHCs,
as well as the properties of their outbursts, and the energy output in the form of a jet.
Our discussion is based on the assumption that the jet
power can be traced by the radio and near-IR normalizations of the radio/X-ray and OIR/X-ray correlations (Fender et al. 2010). However,
some sources might not follow a correlation with slope $b\sim0.6$.
Table \ref{tab:rms_residuals} reports the rms of the residuals obtained from fitting the X-ray/radio and near-IR/X-ray data.
Figure \ref{fig:normalizations_VS_rms} shows the rms of the residuals as a function of the radio and near-IR normalizations.
In the left-hand panel, the BHCs H 1743-322 and GRO J1655-40 are characterized by a much higher rms than all the other
sources (a factor 5.3 and 4.3 higher than the rms of the residuals of the prototypical system GX 339-4). This suggests that a fit
with a slope $b\sim0.6$ is probably not satisfactory and the data points have a high dispersion around the best
fit relation (see Jonker et al. 2010 for H 1743-322). However, other ``radio quiet'' BHCs for which more than 1 data
point (in X-ray/radio) is available are characterized by a rms higher than the one of GX 339-4 only by a factor 1.7 (Swift J1753.5-0127), 
1.5 (XTE J1650-500) or even lower (XTE J1720-318). In the right-hand panel, the ``radio quiet'' BHC XTE J1550-564 has the highest rms,
a factor 2 higher than the second highest, XTE J1118+480. The only other ``radio quiet'' BHC in the plot,
Swift J1753.5-0127, has actually the lowest rms (but there are only 7 sources in the sample). These considerations suggest
that fitting the X-ray/radio and X-ray/near-IR data fixing the slope to the value $b\sim0.6$ constitutes a good method or
at least does not particularly affect the result for the ``radio quiet'' sources.

Figure \ref{fig:powerVSorbital_param} and the Spearman correlation coefficients in Table \ref{tab:log_spearman}
show that there is no connection between two (non independent) parameters of the binary system (the size of the
Roche lobe of the black hole and the orbital period) and the radio and near-IR normalizations.
This result is not unexpected: theoretical works predict that a powerful jet should be formed
when a thick accretion flow is present (as it is thought the case in the LHS; Livio et al. 1999; Meier 2001).
Considering that the thick accretion flow does not extend to the outer regions of the accretion disc (although
the details of its geometry are unknown, see Gilfanov 2009 and references therein), it seems unlikely that the jet power
is affected by the size of the Roche lobe and the orbital period.

We repeated the same analysis using a restricted sample of BHCs for which both the mass and the distance have been
measured (\S \ref{par:binary_restricted}). Our conclusions do not change,
suggesting that our results are probably not strongly biassed by the large uncertainties in the measures
of the mass and the distance.
Our conclusions are unvaried even if we apply the mass correction from the fundamental plane
of black hole activity.

In \S \ref{par:outburst} we examined the spectral properties of the outbursts occurred by the BHCs in our sample.
Our results suggest that the energy output in the jet in the LHS is not regulated by the fact that the
BHC transits or not to the soft states.

It is worth to note that the three of the four known BHCs with the shortest orbital periods
(Swift~J1753.5-0127, XTE~J1118+480 and GRO~J0422+32) only had hard outbursts (but see
Negoro et al. 2009 for a possible softening of Swift~J1753.5-0127 and Brocksopp et al. 2010 for the last outburst of XTE~J1118+480).
Recently, Negoro et al. (2010) discovered a new BHC, MAXI J1659-152. This source might be the BHC with the shortest
orbital period (2 hours and 25 minutes, see Belloni et al. 2010a and Kuulkers et al. 2010). However, this system had a normal outburst,
with a transition to the soft states (see Belloni et al. 2010b and Shaposhnikov \& Yamaoka 2010).
This suggests that a bigger sample of BHCs characterized by a short orbital period will certainly help to clarify
this issue.

From binary evolution calculations, BHCs can in principle have orbital periods as short as $\sim$2 hours and
evolutionary models actually predict that short-period systems might form the majority of them,
similarly to what is observed in cataclysmic variables (Yungelson et al. 2006). The fact that 16 BHCs in
Table \ref{tab:log_sources} have an orbital period $P_{orb} \geq 4.1$ hr is quite puzzling.
A possible explanation is that in short-period systems, the mass transfer from
the companion star might be interrupted by resonances within the primary's Roche lobe, if the mass ratio
is $q \sim 50$ (Yungelson et al. 2006). Zurita et al (2008) suggested that the mass ratio in Swift~J1753.5-0127
is $q \gtrsim 10$ and could be as high as approximately $40$, thus making Swift J1753.5-0.127 the first BHC
detected nearly in this regime. However, even if the mass transfer in Swift J1753.5-0127 is partially
interrupted (because of the high mass ratio), this does not explain why the source is less luminous
than expected in the radio band (in other words why its radio and near-IR normalizations are low compared
to the majority of the BHCs in our sample), although its X-ray luminosity is comparable to
other BHCs (Soleri et al. 2010). The behaviour of XTE~J1118+480 is different: although it is the
known BHC with the second shortest orbital period ($P_{orb} = 4.1$ hr) and its mass ratio is rather
high compared the other sources in the sample ($q \sim 27$), its radio normalization is higher than the one
estimated for 7 other BHCs with longer orbital periods. This suggests that it features jet properties
that are not different from the majority of the BHCs.

\begin{figure}
\begin{tabular}{c}
\resizebox{8.5cm}{!}{\includegraphics{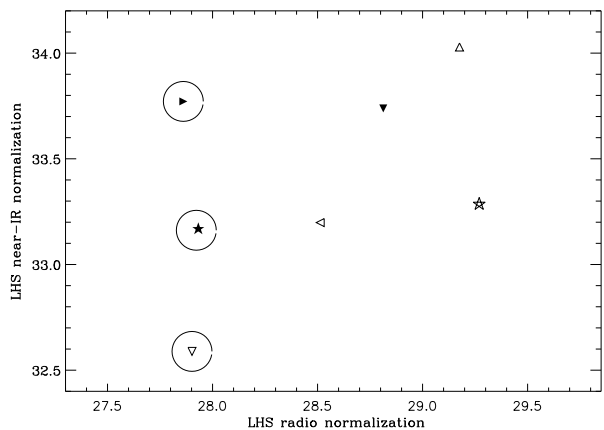}}
\end{tabular}
\caption{Near-IR normalizations versus the radio normalizations. Big circles mark the ``radio quiet'' sources.
See Figure \ref{fig:powerVSorbital_param} for a key to the symbols.}
\label{fig:radio_VS_JHK_normalization}
\end{figure}

In \S \ref{par:binary} we investigated the connection between the inclination angles of the BHCs and the
radio and near-IR normalizations.
The upper panel of Figure \ref{fig:power_outburst} does not show any correlation between the radio normalization
and the inclination angles (see also Table \ref{tab:log_spearman}). The near-IR normalization instead has a fairly high
probability to be anticorrelated to the inclination angle.
This would be consistent with a scenario in which the inner jet (responsible for the IR emission,
Blandford \& K\"{o}nigl 1979) has a higher Lorentz factor $\Gamma$ (and therefore it is more de-boosted)
than the the outer jet (where the radio emission originates), i.e. a decelerating jet. If we restrict our
analysis on a sample of sources for which both the mass and the distance have been
measured (\S \ref{par:binary_restricted}) neither the radio nor the near-IR normalizations are correlated to the inclination
angle. However, this result might have been strongly biassed by the limited number of data points.
The radio jet power does not appear to be correlated to the viewing angle even
if the radio normalizations are calculated using the mass-corrected radio luminosities.
To further investigate the decelerating jet scenario, in Figure \ref{fig:radio_VS_JHK_normalization} we
reported the near-IR normalizations versus the radio normalizations. For a decelerating jet, we do not expect the data to
be linearly correlated, especially for high-inclination sources. Considering all the 7 points, we calculates a Spearman
rank coefficient $\rho \sim 0.3$, with a probability for the null hypothesis of $\sim 43 \%$. If we exclude the data
point for the BHC GRO J1655-40 and XTE J1550-564 (which have very high rms of the residuals in X-ray/radio
and X-ray/near-IR fits, respectively; see Figure \ref{fig:normalizations_VS_rms} left-hand panel) the Spearman rank
correlation coefficient is $\rho \sim 0.7$, with a probability for the null hypothesis of $\sim 16 \%$. If we consider
only high-inclination systems (above $\sim 70^{\circ}$; in this case we exclude the BHCs 4U 1543-47 and GX 339-4) the
Spearman rank correlation coefficient is $\rho \sim 0$, with a probability for the null hypothesis of $\sim 99 \%$.
Despite the number of data points is extremely limited (7 or 5), our results could support the decelerating
jet scenario. More coordinated X-ray, near-IR and radio observations are needed to enlarge the sample.

To further test if Doppler de-boosting effects play a role in populating the sample of BHCs characterized by a faint jet, 
in \S \ref{par:boosting} we defined a jet-toy model.
This model results in a distribution in the ($L_X,L_R$) plane which broadens at higher luminosities.
Considering a range of viewing angles (Figure \ref{fig:toy_model}), the model can qualitatively describe the
scatter around the radio/X-ray correlation. The line for $i \sim 18^{\circ}$ ($\mbox{cos}\, i = 0.95$)
increases its slope at $L_{X} \gtrsim 10^{-4}\, L_{Edd}$ (Figure \ref{fig:toy_model}), since the jet is boosted.
For higher inclination angles, the model predicts that the radio/X-ray correlation, because of de-boosting effect,
should become flatter at high $L_X$. This might be the case for the BHC H~1743-322 (for which $L_R \approx L_X^{\sim0.18}$;
Jonker et al. 2010), in agreement with the high rms of the residuals that we obtained for this source.
Unfortunately, the viewing angle of H~1743-322 system is unknown.
However, the jet-toy model has strong limitations in explaining the behaviour of other systems (e.g. XTE~J1650-500 and
Swift~J1753.5-0127). These ``radio quiet'' BHCs feature a slope of the X-ray/radio correlation steeper than $b \sim 0.6$
(even if Figure \ref{fig:normalizations_VS_rms} seem to suggest that fitting their data point with a slope fixed
to $b \sim 0.6$ is a reasonable approximation),  and they are probably characterized by high inclination angles
($50 \pm 3^{\circ}$ and $> 85 ^{\circ}$, respectively).
We also note that the BHC V404~Cyg lies approximately on the line for $i \sim 18^{\circ}$. This does not match the measured
value of $55 \pm 4^{\circ}$ (Table \ref{tab:log_sources}). The same happens for other BHCs in our sample,
which are actually scattered over several lines.
Our considerations suggest that, although the model can qualitatively describe the scatter in the ($L_X,\, L_R$) plane, it
should be only seen as a viable possibility to describe the ``radio quiet'' population. In fact, at the moment we do not
know whether the ``radio quietness'' is an intrinsic property or it might change with time.

In \S \ref{par:comparison_NS} we compared the ``radio quiet'' BHCs to the NSs. A two-dimensional K-S test can not completely
rule out the possibility that the two families are statistically indistinguishable in the X-ray:radio plane, if a mass
correction is applied. Interestingly, including or not the NS Z sources (that usually feature a disc-jet coupling more similar to BHCs than to NS atoll sources, Migliari \& Fender 2006) does not substantially change the results of the K-S test.
Since for several outliers the mass of the compact object has not been dynamically measured (e.g. GRS 1758-258 and XTE J1720-318),
our results seem to suggest that some of the outliers could actually be neutron stars. Another possibility is that the disc-jet
coupling in the ``radio quiet'' BHCs is more similar to the one in NSs.
\begin{table*}
\centering
\caption{Two-dimensional K-S test for 2 combination of data sets, with or without including the mass correction from the fundamental plane of Merloni et al. (2003). The ``radio quiet'' BHCs are GRS 1758-258, XTE J1550-564, XTE J1650-500, GRO J1655-40, XTE J1720-318 and Swift J1753.5-0127 (see their position in Figure \ref{fig:BH_NS}). In the atoll sources sample  we included the accreting millisecond X-ray pulsars. In the last column we report the probability of the null hypothesis, i.e. the probability that the two data sets are statistically indistinguishable.}
\label{tab:KS}
\begin{tabular}{c c c c c}
\hline
\hline
                              \multicolumn{5}{c}{{\bf Two-dimensional Kolmogorov-Smirnov test}}                             \\
First sample           &    Second sample       &   mass correction        &     K-S statistic (D)   &   Prob (D $>$ observed)        \\	
\hline  
``Radio quiet'' BHCs   &         NSs            &        yes               &      0.46               &       $0.13 \%$                \\
``Radio quiet'' BHCs   &         NSs            &        no                &      0.63               &       $1.9 \times 10^{-4} \%$  \\
``Radio quiet'' BHCs   &    Atoll sources       &        yes               &      0.55               &       $0.02 \%$                \\  
``Radio quiet'' BHCs   &    Atoll sources       &        no                &      0.70               &       $5.0 \times 10^{-5} \%$  \\   
\hline  
\hline
\end{tabular}
\end{table*}

Table \ref{tab:KS} also shows that if a mass correction from the fundamental plane of Merloni et al. (2003) is not applied,
``radio quiet'' BHCs and NSs clearly do not belong to the same population. It is interesting to note that the mass correction
works for black holes of different masses as well as NSs: this implies that the correction may not be dependent on black-hole
features like for example the presence of an event horizon.

\section{Conclusions}
We examined three characteristic parameters of BHCs and the properties of their outbursts to test whether they regulate the energy
output in the form of a jet. This has been motivated by the fact that a growing population of sources seems to feature similar
accretion flows to the majority of the BHCs (e.g. similar radiative efficiency) but fainter
jets than expected. Garcia et al. (2003) suggested that spatially-resolved powerful jets
(so discrete ejections and not compact-steady jets in the LHS) might be associated with long period systems. If our estimates
of the jet power are correct, both the orbital period and the size of the accretion disc are not related to the radio
and near-IR jet power.

We retrieved the properties of the outbursts occurred by the BHCs in our sample to see if LHS-only outbursts feature
different jet properties. We did not find any association between the jet power and the fact that a BHC transits
to the soft states during an outburst.

We also considered the inclination angles for our sample of BHCs.
A recent result shows that compact-steady jets in the LHS might have bulk Lorentz factor $\Gamma >2$. 
This suggests that not only jets with a high inclination can suffer de-boosting effects. However, we did not find any
association between the viewing angles and the jet power inferred from radio observations. The jet power obtained
from near-IR measurements decreases when the inclination angle increases. Although this result might be biassed by the small
number of BHCs for which we have IR data, it could favour a scenario in which the jet decelerates moving from the IR-emitting
to the radio-emitting part.

We defined a jet-toy model in which the bulk Lorentz factor $\Gamma$ becomes larger than $\sim 1$ above $0.1 \% \, L_{Edd}$.
Considering an uniform distribution of viewing angles in the $\mbox{cos} \, i$ space, the model results in a distribution
in the ($L_X,L_R$) plane which broadens at higher luminosities. The model can qualitatively reproduce the scatter around
the radio/X-ray correlation. Although this result is quite promising, we stress that the toy-model has several limitations,
for instance it can not reproduce the slope of the X-ray/radio correlation for some ``radio quiet'' BHCs characterized by
high inclination angles.
Nevertheless we think that it suggests a valid possibility that theoretical models should explore in more detail.

We finally compared the ``radio quiet'' BHCs to the NSs. A two-dimensional K-S test can not completely rule out the
possibility that the two families are statistically indistinguishable in the X-ray:radio plane, if a mass correction
from the fundamental plane of black hole activity is applied. This result suggests that some ``radio quiet'' BHCs could
actually be NSs; alternatively it suggests that some BHCs feature a disc-jet coupling more similar to NSs rather than to
the majority of the BHCs.

\section*{Acknowledgments}
PS acknowledges support from NWO (Netherlands Foundation for Scientific Research).
We thank Elena Gallo and David Russell for providing the original data used to calculate the radio and near-IR normalizations and Michiel van der Klis for very useful comments on the early drafts of the manuscript. The authors also thank Simone Migliari for providing the neutron star data
and the referee for giving very useful comments that certainly improved the quality of this manuscript. PS would also like to thank
Piergiorgio Casella, Alessandro Patruno, Tomaso Belloni, Andrea Sanna, Mariano M{\'e}ndez and Monica Colpi for useful discussion.


\label{lastpage}

\end{document}